\documentclass[12pt,draftcls,onecolumn]{IEEEtran}
\usepackage{graphicx}

\usepackage{amsmath,amsfonts,amssymb,amsthm}
\usepackage{latexsym}
\usepackage{bm}
\usepackage{color}

\newtheorem{definition}{Definition}
\newtheorem{assumption}{Assumption}

\newtheorem{theorem}{Theorem}

\title{Rate-Distortion Analysis of Quantizers with Error Feedback}

\author{Shuichi Ohno$^{\ast}$, Teruyuki Shiraki, M. Rizwan Tariq, and Masaaki Nagahara\\
$^{\ast}$Hiroshima University, 
1-4-1 Kagamiyama, Higashi-Hiroshima 739-8527, JAPAN\\
}

\begin{document}

\maketitle

\begin{abstract}
A $\Delta\Sigma$ modulator that is often utilized to convert analog signals into digital signals can be modeled as a static uniform quantizer
with an error feedback filter. 
In this paper, we present a rate-distortion analysis 
of quantizers with error feedback including the $\Delta\Sigma$ modulators,  
assuming that the error owing to overloading in the static quantizer 
is negligible. 
We demonstrate that the amplitude response of 
the optimal error feedback filter that minimizes 
the mean squared quantization error can be parameterized by one parameter. 
This parameterization enables us to determine the optimal error feedback filter numerically. 
The relationship between the number of bits used for the quantization and 
the achievable mean squared error can be obtained  
using the optimal error feedback filter. 
This clarifies the rate-distortion property of quantizers with error feedback. 
Then, ideal optimal error feedback filters are approximated 
by practical filters using the Yule-Walker method 
and the linear matrix inequality-based method.    
Numerical examples are provided for demonstrating our analysis and synthesis. 
\end{abstract}


\clearpage
\section{Introduction}
\label{sec:introduction}
Quantization is a fundamental process in digital processing,
wherein, 
a large set of input values are mapped onto a smaller set of output values. 
Analog signals have to be quantized into digital signals. 
The simplest type of quantizer is the uniform quantizer  
that has fixed-length code words, i.e., a fixed number of bits per sample. 
However, the uniform quantizer is not efficient 
because it does not consider 
the statistics of the input and/or the information about 
the system connected to the quantizer.  
Additional information regarding the input and/or the connected system  
can be exploited to obtain good quantizers.  
Under the assumption that the quantization error is 
a white uniformly distributed random sequence, 
the Lloyd-Max quantizer is optimal 
among the quantizers having fixed-length code words  
in the sense that it minimizes the distortion of the quantization error 
\cite[Chap.9]{sayood2012introduction}. 
However, the probability density function of the input to the quantizer, 
that is often unavailable in practice, 
is required for constructing the Lloyd-Max quantizer. 

Quantization with error feedback is more efficient than  
the conventional uniform quantization.  
It includes a uniform quantizer and a feedback filter, where 
the filtered error of the uniform quantizer 
is fed back to it 
for mitigating the error introduced by quantization. 
Quantization with error feedback is used 
for reducing the effect of the quantized coefficients 
in fixed-point digital filters \cite{1084254,134473}. 
Finite impulse response (FIR) error feedback filters  
have been proposed for recursive digital filters composed 
of cascaded second order sections in \cite{1163989}. 

Various designs for the feedback filter have been proposed. 
Based on the generalized Kalman-Yakubovich-Popov (GKYP) lemma, 
an FIR error feedback filter has been designed 
to minimize the worst case gain in the signal passband   
using convex optimization \cite{6156470}, 
whereas an infinite impulse response (IIR) filter 
using an iterative algorithm \cite{li2014design}. 
Under the whiteness assumption for the error of the uniform quantizer, 
an optimal FIR feedback filter that minimizes the variance of the error 
owing to quantization has been proposed in \cite{6461108}. 
On the other hand, IIR error feedback filters have been presented 
in \cite{ohno2015optimal} for minimizing the maximum absolute value  
of the error in the signal of interest, introduced by the quantization.  

Quantization with error feedback is also adopted in  
$\Delta\Sigma$ or $\Sigma\Delta$ modulators 
that are often utilized to convert 
real values into fixed-point numbers and vise versa \cite{Delta-Sigma}.  
$\Delta\Sigma$ modulators are widely used for several applications, e.g.,  
audio signal processing \cite{1226728}, 
RF transmitter architectures \cite{5518336}, 
compressive sensing \cite{boufounos20081}, 
and independent source separation \cite{5290030}.

It is known that when a $\Delta\Sigma$ modulator is used 
to quantize an analog signal 
into a digital signal, oversampling can effectively 
reduce the error introduced by quantization. 
However, oversampling increases the number of bits per time, 
if the same number of bits are assigned to each output of the quantizer. 
Whether oversampling is effective when the number of bits per time is fixed, continues to remain unclear. 
To determine this, the rate-distortion analysis of 
the $\Delta\Sigma$ modulator is necessary. 

It has been found in \cite{492534} that for bandlimited signals,  
the variance of the distortion, i.e., the mean squared error (MSE)   
of a simple single-loop one-bit $\Delta\Sigma$ modulator decays 
at a rate of $O(\lambda^{-4})$, where $\lambda$ is the oversampling ratio. 
In \cite{daubechies2003approximating}, 
it is proven that for bandlimited bounded signals  
the squared maximum absolute value, i.e., 
the squared $l_{\infty}$ norm of 
the distortion of a one-bit $\Delta\Sigma$ modulator 
can decrease at a rate of $O(\lambda^{-4})$; 
further, a family of one-bit $\Delta\Sigma$ modulators that attain 
this rate has been provided. 
In \cite{deift2011optimal}, 
optimal filters in this family are designed to minimize 
the decay rate demonstrating that 
an exponential rate of $O(2^{-r\lambda})$ for $r\approx 0.102$ 
is achieved by the designed filter. 
On the other hand, for bandlimited stationary signals, 
the MSE of the optimal one-bit $\Delta\Sigma$ modulator 
that minimizes the MSE under a constraint on 
the variance of the input to the uniform quantizer, 
decreases exponentially at a rate of $O(2^{-r\lambda})$ 
for $r\approx 0.807$ \cite{4522532}. 
This improvement becomes possible by exploiting the knowledge on 
the power spectral density function  
of the input that is not always available 
and by using an additional pre-filter and post-filer with an infinite order.  
In this paper, we consider a more practical situation, wherein  
the spectrum of the input is unavailable, 
and clarify the rate-distortion relationship  
of the conventional $\Delta\Sigma$ modulators without pre-/post-filters.  

The input to the static quantizer in a quantizer with an error 
feedback exhibits a larger amplitude than the input to 
a conventional uniform quantizer without an error feedback.  
To enable fair comparisons between quantizers with different 
input amplitudes, we assume that the error variance of the static quantizer 
is proportional to the variance of its input. 
Under this assumption, we study the variance of the error 
at the output of the system connected to the quantizer. 
This enables a rate-distortion analysis of quantizers with error feedback. 

After formulating our problem as an optimization problem, 
we show that the amplitude response of 
the optimal error feedback filter that minimizes the MSE at the output 
can be parameterized by one parameter. 
Then, the optimal error feedback filter can be determined numerically 
by minimizing the MSE with respect to the parameter. 
The relationship between the number of bits used for quantization and 
the achievable MSE can be revealed. 
This is our main contribution on the rate-distortion analysis 
of quantizers with error feedback. 
It guarantees that if a fixed number of bits are assigned 
for the quantization, the optimal quantizer with an error feedback 
outperforms the uniform quantizer. 
It also demonstrates the contribution of oversampling to the reduction of the MSE. 
Finally, we develop two approximations for ideal optimal filters  
using the Yule-Walker method \cite{4103917} and 
the linear matrix inequality-based method 
for obtaining practical error feedback filters.  
Numerical examples are provided to demonstrate our analysis and synthesis. 


\section{Quantizer with error feedback}
\label{sec:quantizer}

\begin{figure}[tbp]\setlength{\unitlength}{10mm}
  \centering
 \begin{center}
\leavevmode
\begin{picture}(7,3.5)(0,-0.5)
\put(0,2){\vector(1,0){0.9}}
\put(1,2){\circle{0.2}}
\put(1,2){\makebox(0,0){\small +}}
\put(1.1,2){\vector(1,0){3.4}}
\put(4,2){\circle*{0.1}}
\put(4.5,1.5){\framebox(1,1){$Q(\cdot)$}}
\put(5.5,2){\vector(1,0){1.5}}
\put(6,2){\circle*{0.1}}
\put(4,2){\line(0,-1){1}}
\put(6,2){\line(0,-1){1}}
\put(4,1){\vector(1,0){0.9}}
\put(6,1){\vector(-1,0){0.9}}
\put(5,1){\circle{0.2}}
\put(5,0.9){\line(0,-1){0.9}}
\put(5,0){\vector(-1,0){1}}
\put(2,-0.5){\framebox(2,1){$R[\mathrm z]-1$}}
\put(2,0){\line(-1,0){1}}
\put(1,0){\vector(0,1){1.9}}
\put(4.5,-0.3){\makebox(0,0)[c]{$w$}}
\put(0.5,2.3){\makebox(0,0)[c]{$x$}}
\put(4,2.3){\makebox(0,0)[c]{$u$}}
\put(6,2.3){\makebox(0,0)[c]{$v$}}
\put(5.2,0.7){\makebox(0,0)[c]{\small $+$}}
\put(4.7,0.7){\makebox(0,0)[c]{\small $-$}}
\end{picture}
\end{center}
\caption{Quantizer with an error feedback filter}
\label{fig:efb_ruantizer_c}
\end{figure}
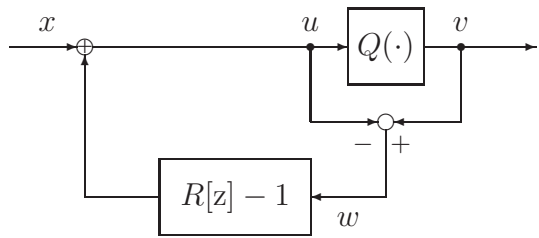

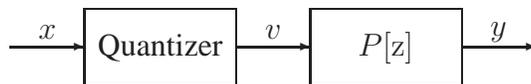
\begin{figure}[tbp]\setlength{\unitlength}{10mm}
  \centering
 \begin{center}
\leavevmode
\begin{picture}(7,3)(0,0)
\put(0,1){\vector(1,0){1}}
\put(1,0.5){\framebox(2,1){Quantizer}}
\put(3,1){\vector(1,0){1}}
\put(4,0.5){\framebox(2,1){$P[\mathrm z]$}}
\put(6,1){\vector(1,0){1}}
\put(0.5,1.2){\makebox(0,0)[c]{$x$}}
\put(3.5,1.2){\makebox(0,0)[c]{$v$}}
\put(6.5,1.2){\makebox(0,0)[c]{$y$}}
\end{picture}
\end{center}
\caption{Quantizer and system}
\label{fig:QandS}
\end{figure}

Figure \ref{fig:efb_ruantizer_c} depicts our quantizer with the error feedback, where $x$ is the input signal to the quantizer with the error feedback, $v$ is its output signal, and $Q(\cdot)$ denotes a conventional 
static uniform quantizer. 
All the signals are assumed to be of discrete-time. 
We denote the $\mathrm z$ transform of a discrete-time signal, 
$f=\{f_k\}_{k=0}^{\infty}$, as $F[z]=\sum_{k=0}^{\infty}f_k \mathrm z^{-k}$. 
We also express the output signal $b$ 
of the linear time invariant (LTI) system, 
whose transfer function is $F[z]$,   
to the input $a=\{a_k\}_{k=0}^{\infty}$ as $b=F[z] a$, 
where $z^{-1}$ is a unit-time delay operator. 

In Fig. \ref{fig:efb_ruantizer_c}, 
the signal $w=v-u$ is the quantization error signal of 
the static uniform quantizer that is filtered by $R[\mathrm z]-1$ 
and fed back to $x$. 
The first coefficient of the impulse response of $R[\mathrm z]$ is 
assumed to be one, implying that $R[\mathrm z]-1$ is strictly causal 
and hence, practically implementable. 
The minus one in $R[\mathrm z]-1$ is only for the simplicity of presentation. 

Quantization with error feedback has a simple structure  
that can be implemented at a relatively low cost. 
The linearized model of the $\Delta\Sigma$ modulator can be  
expressed by a quantizer with an error feedback filter.  

The input signal $u$ to the uniform quantizer is expressed as 
$u=x+(R[\mathrm z]-1)w$. 
The quantization error signal of the quantization with the error feedback can 
be defined as $e=v-x$ that should be differentiated with 
the quantization error signal $w$ of the uniform quantizer.   
It is easily discernible that they are related through $e=R[\mathrm z]w$. 
Then, the output of the quantizer can be expressed as  
\begin{equation}
  \label{eq:1}
  v=x+R[\mathrm z]w. 
\end{equation}
From \eqref{eq:1}, the effect of the quantization noise $w$ can be reduced 
by the filter $R[\mathrm z]$.  
Quantization with an error feedback has been used to mitigate 
quantization errors in digital filters as well as in $\Delta\Sigma$ modulators. 
As $R[\mathrm z]$ shapes the spectrum of the noise $w$, it is 
called a {\it noise shaping filter} or a {\it noise transfer function}. 
For $\Delta\Sigma$ modulators, $R[\mathrm z]$ 
has been designed to minimize the maximum of 
the amplitude response $|R[e^{j\omega}]|$ in the passband of $x$ 
\cite{6156470,li2014design}.  

We assume that the output of the quantizer passes through 
the system $P[\mathrm z]$ as depicted in Fig. \ref{fig:QandS}. 
The output $y$ of $P[\mathrm z]$ can be expressed as 
$y=P[\mathrm z]v=P[\mathrm z]x+\epsilon$,   
where $\epsilon$ is the error at the output 
introduced by the quantization and is given by  
\begin{equation}
  \label{eq:2}
\epsilon=P[\mathrm z]R[\mathrm z]w.   
\end{equation}

If we know the statistics of the input $x$ and/or 
the system $P[\mathrm z]$ connected to the quantizer, 
we can design the noise shaping filter. 
For example, the bandlimitedness of the input is utilized in 
\cite{492534,daubechies2003approximating,deift2011optimal}, 
whereas $P[\mathrm z]$ is exploited in \cite{6156470, li2014design,6461108}. 
Both the input spectrum and $P[\mathrm z]$ 
are taken into account in \cite{4522532}, where  
the input, $x$ and the output, $v$ 
are processed by a pre-filter and a post-filter, respectively, 
that are dependent on the input spectrum and $P[\mathrm z]$. 
It has been shown \cite{4522532} that 
the rate of the optimal one-bit $\Delta\Sigma$ modulator 
decreases exponentially at a rate of $O(2^{-r\lambda})$ for $r\approx 0.807$. 
However, the input spectrum is often unavailable in practice.  
The purpose of this paper is to clarify 
the rate-distortion relationship of the quantizer with an error feedback, 
when the input spectrum cannot be used. 

\section{Optimal Noise Shaping Filter}
\label{sec:optimal}

First, let us review static uniform quantizers. 
Although most of our analysis holds true 
for the other types of static quantizers under the same conditions, 
we consider the mid-rise quantizer as an example.   

The mid-rise quantizer can be described by two parameters,
the quantization interval $d(>0)$ and the saturation level $L(>0)$. 
Its output for a scalar input $\xi$ is expressed as 
\begin{equation}
  \label{eq:3}
  Q(\xi)=
\left\{
  \begin{array}{cc}
    \left(
      i+\frac{1}{2}
    \right)     d, &\xi \in [id, (i+1)d)  \\
    &  \mbox{for an integer $i$ and} \ \ |\xi|\leq L+\frac{d}{2} \\
    L,& \xi >  L +\frac{d}{2}\\
    -L, &\xi <  -L-\frac{d}{2} \\
  \end{array}
\right.
. 
\end{equation}
The overload is the saturation owing to the fixed number of bits 
representing the binary-values. 
In the mid-rise quantizer, 
an overload occurs if $|\xi| > L+\frac{d}{2}$. 

If we assign $b$ bits to the mid-rise quantizer, where 
$b$ is a positive integer, 
the number of quantization levels is $2^b$ that is 
related to the dynamic range $[-L,L]$ of the mid-rise quantizer 
and the quantization interval $d$ through  
\begin{equation}
  \label{eq:4}
  2L=(2^b-1) d.
\end{equation}

For our analysis, as in \cite{4522532}, 
we assume that a sufficient number of bits 
are assigned to the output of the uniform quantizer so that:  
\begin{assumption}\label{assumption:0}
  The error owing to the overload is negligible. 
\end{assumption}
The input $x$ to our quantizer is assumed to be a wide-sense stationary 
process with a zero mean and a variance $\sigma_x^2$.   
We also assume that 
the quantization error signal of the static uniform quantizer is a white noise 
and is uncorrelated with the input $x$. 
\begin{assumption}\label{assumption:1}
  The quantization error signal $w$ of the uniform quantizer 
  is a white random signal with a zero mean and a variance $\sigma_w^2$ 
  and uncorrelated with the input of the uniform quantizer. 
\end{assumption}

The dynamic range of the static quantizer is determined 
by the dynamic range of its input. 
It is reasonable to assume that the dynamic range of the static quantizer 
is proportional to the dynamic range of the amplitude of its input,  
when the number of bits assigned to the uniform quantizer is fixed.  
Under Assumption \ref{assumption:1}, 
we assume as in \cite{tuqan1997statistically} that: 
\begin{assumption}\label{assumption:2}
  For a fixed number of quantization levels, the variance of 
the quantization error of the uniform quantizer 
is proportional to the variance of its input 
and the ratio is defined as 
\begin{equation}
  \label{eq:5}
  \gamma=\frac{\sigma_u^2}{\sigma_w^2}
\end{equation}
where $\sigma_u^2$ and $\sigma_w^2$ are 
the variances of the input and the quantization error, respectively.  
\end{assumption}
This assumption enables us to analyze quantizers 
with different dynamic ranges.  

Let us denote the $L_2$ norm of a filter $H[\mathrm z]$ 
as $||H[\mathrm z]||$ that is defined as  
\begin{equation}
  \label{eq:6}
  ||H[\mathrm z]||=
  \left(
\frac{1}{2\pi} \int_{-\pi}^{\pi}
  H^{\ast}[e^{j\omega}]H[e^{j\omega}] d\omega
  \right)^{\frac{1}{2}}
\end{equation}
where $c^{\ast}$ is the complex conjugate of $c$. 

From Assumption \ref{assumption:1}, the variance of 
the input $u$ to the uniform quantizer is expressed as 
\begin{equation}
  \label{eq:7}
  \sigma_u^2=\sigma_x^2+||R[\mathrm z]-1||^2\sigma_w^2
.
\end{equation}
Then, under Assumption \ref{assumption:2}, 
the variance of the quantization error of the uniform quantizer is 
expressed as 
\begin{equation}
  \label{eq:8}
  \sigma_w^2=\frac{\sigma_x^2}{\gamma-||R[\mathrm z]-1||^2}, 
\end{equation}
that requires 
\begin{equation}
  \label{eq:9}
  \gamma-||R[\mathrm z]-1||^2>0.
\end{equation}
This implies that the energy of the feedback signal has to be limited. 
As the first entry of the impulse response of $R[\mathrm z]$ is unity, 
we have $||R[\mathrm z]-1||^2+1 = ||R[\mathrm z]||^2$ and then  
\begin{equation}
  \label{eq:10}
  \sigma_w^2=\frac{\sigma_x^2}{\gamma+1-||R[\mathrm z]||^2}.
\end{equation}

The variance of the quantization error 
at the output of the system is obtained from \eqref{eq:2} by 
\begin{equation}
  \label{eq:11}
  ||P[\mathrm z]R[\mathrm z]||^2\sigma_w^2. 
\end{equation}
Substituting \eqref{eq:10} in \eqref{eq:11} results in 
\begin{equation}
  \label{eq:12}
  ||P[\mathrm z]R[\mathrm z]||^2\sigma_w^2
=
\frac{||P[\mathrm z]R[\mathrm z]||^2}{\gamma+1-||R[\mathrm z]||^2}
\sigma_x^2
.
\end{equation}

To observe the performance of our quantizer, 
we would like to obtain the optimal noise shaping filter  
$R[\mathrm z]$ and the minimum of the mean squared error (MSE).  
For a given $\sigma_x^2$ and $P[\mathrm z]$, 
we minimize the MSE with respect to $R[\mathrm z]$. 
To stabilize the quantizer, $R[\mathrm z]$ must be stable.  
Then, as $\sigma_x^2$ in \eqref{eq:12} is a constant,  
our problem can be formulated as the following minimization:
\begin{equation}
  \label{eq:13}
  \min_{R[\mathrm z] \in RH_{\infty}} 
\frac{||P[\mathrm z]R[\mathrm z]||^2}{\gamma+1-||R[\mathrm z]||^2}
\end{equation}
subject to $R[\infty]=1$ and 
\begin{equation}
  \label{eq:14}
  ||R[\mathrm z]||^2<\gamma+1
\end{equation}
where $RH_{\infty}$ is the set of 
stable proper rational functions with real coefficients.

To enable theoretical analysis, 
we relax the stable proper rational function $R[\mathrm z]$ to 
a function  $r(\omega) \in L_2$  
belonging to a more general class of functions 
that is piece-wise differentiable on $[-\pi,\pi]$,    
has at most a finite number of discontinuity points,    
and satisfies 
\begin{align}
\label{eq:15}
& \frac{1}{2\pi} \int_{-\pi}^{\pi} \ln r(\omega) d \omega =c_0 
\end{align}
for $c_0 \geq 0$. 
We note that \eqref{eq:15} 
is imposed by the stability of the original function $R[\mathrm z]$. 

The $L_2$ norm of $q(\omega) \in L_2$ is defined as
\begin{equation}
  \label{eq:16}
  ||q(\omega)||
=
\left(
\frac{1}{2\pi}\int_{-\pi}^{\pi} q^{\ast}(\omega) q(\omega) d \omega
\right)^{\frac{1}{2}}. 
\end{equation}
We denote the set of $L_2$ functions that satisfy \eqref{eq:15} 
as ${\cal C}_0$. 
We also define a set ${\cal C}_1$ of $L_2$ functions as
\begin{equation}
  \label{eq:17}
  {\cal C}_1=
  \left\{
    r(\omega) : ||r(\omega)||^2 < \gamma+1
  \right\}
.
\end{equation}
Then, we would like to determine $r(\omega) \in {\cal C}_0 \cap {\cal C}_1$ 
that minimizes  
\begin{equation}
  \label{eq:18}
    \frac{||p(\omega)r(\omega)||^2}{\gamma+1-||r(\omega)||^2}
\end{equation}
where 
\begin{equation}
  \label{eq:19}
  p(\omega)= |P[e^{j\omega}]|
.
\end{equation}

Although we extend the class of functions, 
from Lemma 1 in \cite{4522532}, 
we can find a stable proper rational function  
$R[\mathrm z]$ such that $|R[e^{j\omega}]|$ 
approximates $r(\omega)$ arbitrarily well in $[-\pi,\pi]$. 
Then, the stable proper rational function that approximates 
the solution for the minimization of \eqref{eq:18} 
can be considered as an approximate solution 
for the original minimization problem. 

Now, our problem is to find the optimal function that minimizes  
\eqref{eq:18} such that 
\begin{equation}
  \label{eq:21}
  r_{opt}(\omega)=\arg \min_{r (\omega)\in {\cal C}_0 \cap {\cal C}_1} 
\frac{||p(\omega)r(\omega)||^2}{\gamma+1-||r(\omega)||^2}. 
\end{equation}
For our analysis, let us introduce the notion of almost constant functions. 
\begin{definition}
  A function $\psi$ : $[a,b] \rightarrow \mathbb{R}$  
is said to be almost constant if and only if  
\begin{equation}
  \label{eq:20}
  \int_{a}^{b}
  \left|
\psi(x)-\frac{1}{b-a}\int_{a}^{b}\psi(x) dx 
  \right| \psi(x)dx =0
\end{equation}
\end{definition}

The optimal solution for our problem 
cannot be expressed in a closed-form 
but can be characterized with one parameter as follows  
(see Appendix \ref{sec:appendix1} for proof):
\begin{theorem}\label{theorem:1}
Suppose that $p(\omega)$ is not almost constant. 
Then, for any $\gamma>0$, the optimal function 
that minimizes \eqref{eq:18} can be expressed using
a parameter $\alpha$ as 
  \begin{equation}
    \label{eq:22}
    r_{\alpha}(\omega)=\frac{\theta(\alpha)}{\sqrt{p^2(\omega)+\alpha}}
  \end{equation}
where
\begin{equation}
  \label{eq:23}
  \theta(\alpha)=\exp
  \left(
\frac{1}{4\pi}\int_{-\pi}^{\pi} \ln (p^2(\omega)+\alpha) d \omega
  \right)
.
\end{equation}
If $p(\omega)$ is almost constant, then the optimal function is 
almost constant. 
\end{theorem}
It has been shown in \cite{6771600} that 
the optimal noise shaping filter $R[\mathrm z]$ that minimizes 
$||P[\mathrm z]R[\mathrm z]||^2\sigma_w^2$ without any constraint 
on the input to the static quantizer has 
an amplitude response proportional to $1/p(\omega)$. 
Theorem \ref{theorem:1} reveals that the optimal 
noise shaping filter under constraint \eqref{eq:5} has 
a similar amplitude response as the optimal noise shaping filter. 
More importantly, Theorem \ref{theorem:1} assures 
that a quantizer with an error feedback outperforms 
a static uniform quantizer, except for the trivial case 
where $P[\mathrm z]$ is almost constant. 

To proceed further, 
we express our objective function by the parameter $\alpha$ as 
\begin{equation}
  \label{eq:24}
  \Phi(\alpha)=\frac{N(\alpha)}{\gamma+1-C(\alpha)}
\end{equation}
where
\begin{equation}
  \label{eq:25}
  N(\alpha)=\frac{1}{2\pi}
  \int_{-\pi}^{\pi} p^2(\omega) r_{\alpha}^2(\omega) d \omega
\end{equation}
and 
\begin{equation}
  \label{eq:26}
  C(\alpha)=||r_{\alpha}||^2=
\frac{\theta^2(\alpha)}{2\pi}
  \int_{-\pi}^{\pi} 
\frac{1}{p^2(\omega)+\alpha} d \omega
.
\end{equation}

We have to determine the global minimizer of $\Phi(\alpha)$, i.e., 
\begin{equation}
\label{eq:27}
  \alpha_{opt}=\arg \min_{\alpha} \Phi(\alpha).
\end{equation}
In Appendix \ref{sec:appendix2}, we show that 
minimizing $\Phi(\alpha)$ with respect to $\alpha$ 
leads to the following theorem, enabling us to compute 
the minimizer numerically. 
\begin{theorem}\label{theorem:2}
  For any $\gamma>0$, the optimal $\alpha$ 
denoted by $\alpha_{opt}$ that minimizes $\Phi(\alpha)$ satisfies 
$\alpha_{opt}>0$ and 
  \begin{equation}
    \label{eq:28}
    \gamma+1=\frac{\theta^2(\alpha_{opt})}{\alpha_{opt}}
.
  \end{equation}
\end{theorem}

It can be easily discerned that 
\begin{equation}
  \label{eq:29}
  \frac{d}{d \alpha}
  \left(
    \frac{\theta^2(\alpha)}{\alpha}
  \right)
=-\frac{\theta^2(\alpha)}{\alpha^2}
\left(
\frac{1}{2\pi}
  \int_{-\pi}^{\pi} 
\frac{p^2(\omega)}{p^2(\omega)+\alpha} d \omega
\right) <0.
\end{equation}
As $\theta^2(\alpha)/\alpha$ is a monotonically decreasing 
function in $\alpha$,  
$\alpha_{opt}$ that satisfies \eqref{eq:28} 
for a given $\gamma$ can be numerically determined by e.g., the bisection algorithm. 
Then, the optimal function is given by  
\begin{equation}
  \label{eq:30}
      r_{opt}(\omega)=\frac{\theta(\alpha_{opt})}{\sqrt{p^2(\omega)+\alpha_{opt}}}
.
\end{equation}

\section{Rate-Distortion Analysis}
\label{sec:rate-distortion}

Based on the results of the previous section, 
we reveal the relationship between the rate and the distortion 
of the optimal quantizer. 

Let us consider a continuous-time system $P(s)$  
assumed to be bandlimited as follows:
\begin{assumption}\label{assumption:4}
 The continuous-time system $P(s)$, is band-limited in $[-\pi/T_s,\pi/T_s]$ 
  and $1/T_s$ is its Nyquist frequency. 
\end{assumption} 
Under Assumption \ref{assumption:4}, 
it suffices to sample the output of 
the continuous-time system $P(s)$ at the Nyquist rate 
to reconstruct the continuous-time output 
from its sampled discretized output. 

Sampling with a sampling period $T_s/\lambda$ 
when $\lambda$ is a positive integer and $\lambda>1$ 
is known as oversampling. 
The integer $\lambda$ is called the {\it oversampling ratio} and is 
the sampling frequency divided by the Nyquist frequency.

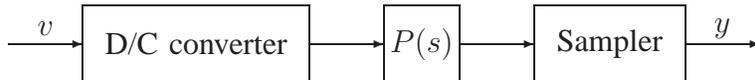
\begin{figure}[tbp]\setlength{\unitlength}{10mm}
  \centering
 \begin{center}
\leavevmode
\begin{picture}(10,3)(0,0)
\put(0,1){\vector(1,0){1}}
\put(1,0.5){\framebox(3,1){D/C converter}}
\put(4,1){\vector(1,0){1}}
\put(5,0.5){\framebox(1,1){$P(s)$}}
\put(6,1){\vector(1,0){1}}
\put(7,0.5){\framebox(2,1){Sampler}}
\put(9,1){\vector(1,0){1}}
\put(0.5,1.2){\makebox(0,0)[c]{$v$}}
\put(9.5,1.2){\makebox(0,0)[c]{$y$}}
\end{picture}
\end{center}
\caption{D/C converter and sampling}
\label{fig:DCS}
\end{figure}

Let us suppose that the output of the quantizer is converted 
by a discrete-time-to-continuous-time (D/C) converter into 
a continuous-time signal and the continuous-time signal passes through 
the continuous-time system $P(s)$ as shown in Fig. \ref{fig:DCS}. 
We sample the continuous-time output signal of $P(s)$ 
to obtain a discrete-time signal $y$. 
Let us denote the discrete-time equivalent system 
from $v$ to $y$ with a sampling period $T_s/\lambda$ as $P_{\lambda}[\mathrm z]$. 

If we utilize an ideal sinc function for our D/C converter such that  
\begin{equation}
  \label{eq:32}
  v(t)=\sum_{k=-\infty}^{\infty} \frac{\sin(\pi(t-kT_s)/T_s)}{\pi(t-kT_s)/T_s}
v_k
\end{equation}
where $v_k$ is the value of the discrete-time signal $v$ at time $k$ 
and $v(t)$ is the reconstructed continuous-time signal. 
Then under Assumption \ref{assumption:4}, the sampled system 
with a sampling period $T_s/\lambda$ satisfies 
\begin{equation}
  \label{eq:33}
  P_{\lambda}[e^{j\omega}]=
  P
  \left(
    \frac{{\lambda}\omega}{T_s}
  \right)
\quad \mbox{for} \quad |\omega| \leq \omega_c. 
\end{equation}

To analyze the relationship between the rate and the distortion of 
the optimal quantizer, we define  
\begin{equation}
  \label{eq:34}
  p_{\lambda}(\omega)=
  \left\{
    \begin{array}{cc }
      p(\lambda\omega) & |\omega|\leq \omega_c\\
      0 & \omega_c < |\omega|\leq \pi \\
    \end{array}
  \right.
\end{equation}
and consider the following minimization problem. 
\begin{equation}
  \label{eq:35}
  \min_{r(\omega) \in {\cal C}_0 \cap {\cal C}_1} 
    \frac{||p_{\lambda}(\omega)r(\omega)||^2}{\nu-||r(\omega)||^2}
\end{equation}
where 
\begin{equation}
  \label{eq:36}
  \nu=\gamma+1
.
\end{equation}
This gives the minimum MSE, or equivalently, the distortion 
of the optimal quantizer. 

Let us denote the minimum of \eqref{eq:35} as $ D(\nu,\lambda)$ that is 
a function in $\nu$ and $\lambda$. 
To designate the dependency of $\alpha_{opt}$ on $\nu$ and $\lambda$, 
we also denote $\alpha_{opt}$ as $\alpha_{opt}(\nu,\lambda)$.  
Substituting \eqref{eq:30} in \eqref{eq:35} and 
using \eqref{eq:28} and \eqref{eq:57}, we find 
\begin{equation}
  \label{eq:37}
  D(\nu,\lambda)=\alpha_{opt}(\nu,\lambda).  
\end{equation}
Using \eqref{eq:37}, we prove in Appendix \ref{sec:appendix3} that: 
\begin{theorem}\label{theorem:3}
Let the oversampling rates be $\lambda$ and $\nu=\gamma+1$, 
where $\gamma$ is defined in Assumption \ref{assumption:2}. 
The MSE of the optimal quantizer with an error feedback 
is a function of $\nu$ and $\lambda$ 
that satisfies  
  \begin{equation}
    \label{eq:38}
    D(\nu,\lambda)=
    D(\nu^\lambda,1)
.
  \end{equation}
\end{theorem}
Let us assume that the uniform quantizer has 
$N=2^b$ quantization levels and an interval of $d$. 
The loading factor is defined as $L_f=L/\sigma_u=Nd/(2\sigma_u)$ 
\cite{gersho2012vector} and is the ratio between $L$ and the 
standard deviation of the input to the uniform quantizer.  
The loading factor regulates the frequency of the overloading. 
For example, if the input to the uniform quantizer is Gaussian, 
then the probability of the input exceeding the range is 
approximately 0.045, when the loading factor is four.  

As the static uniform quantizer cannot 
outperform the quantizer with an error feedback, 
we have 
\begin{equation}
  \label{eq:39}
  D(\nu,1)\leq 
\frac{||P[\mathrm z]||^2}{\gamma}  
=\frac{||P[\mathrm z]||^2}{\nu-1}.  
\end{equation}
It follows from \eqref{eq:38} and \eqref{eq:39} that:  
\begin{theorem}\label{theorem:4}
The MSE of the optimal modulator is upper bounded as  
  \begin{equation}
    \label{eq:40}
    D(\nu,\lambda) \leq 
    \left(
\frac{1}{\nu^{\lambda}-1}
    \right)
 ||P[\mathrm z]||^2.
  \end{equation}
\end{theorem}
Theorem \ref{theorem:4} shows that 
the MSE of the $\Delta\Sigma$ modulator decays 
at a rate of $O(\nu^{-\lambda})$. 
On the other hand, the decay rate of the $\Delta\Sigma$ modulator 
having pre/post-filters and designed with a knowledge of the input spectrum 
is $O(\nu^{-\lambda}/\lambda )$ \cite[Theorem 6]{4522532}  
\footnote{In \cite{4522532}, the decay rate is given by 
$O(\nu^{-\lambda})$, as $p_{\lambda}(\omega)$ in \eqref{eq:34} 
is scaled by $\sqrt{\lambda}$.}.
Thus, we can conclude that 
the term $1/\lambda$ not in $O(\nu^{-\lambda})$ is 
the price we have to pay for the unavailability of the input spectrum.

\section{Design of the Noise Shaping Filters}
\label{sec:design}
We only know the amplitude response of the optimal noise shaping filter 
from the results in Section \ref{sec:quantizer}. 
In practice, we have to implement a noise shaping filter 
with a stable rational transfer function. 
This necessitates the acquisition of an implementable filter
approximating the optimal noise shaping filter. 
 
For approximating a given spectrum, 
the Yule-Walker method \cite{4103917} is well-known, efficient, 
and is optimal in the least squares sense.  
If we permit the usage of a filter with a sufficiently high order, 
then the amplitude response of the approximated filter can be 
almost the same as the amplitude response of the ideal optimal filter. 
However, the head of the impulse response of the noise shaping filter 
has to be unity and this is not assured by the Yule-Walker method in general. 
Although we may be able to modify the Yule-Walker method, 
we only normalize the approximated filter to have a unity 
head for its impulse response. 

Let us develop another approximation to obtain a noise shaping 
filter with a low order. Once 
the amplitude response of the optimal noise shaping is obtained, 
we can compute its norms that are denoted by $||R_{opt}[\mathrm z]||$. 
Then, we consider the following optimization problem:  
\begin{equation}
  \label{eq:41}
  \min_{R[\mathrm z] \in RH_{\infty}} 
||P[\mathrm z]R[\mathrm z]||^2
\end{equation}
subject to $R[\infty]=1$ and 
\begin{equation}
  \label{eq:42}
  ||R[\mathrm z]||^2\leq ||R_{opt}[\mathrm z]||^2.
\end{equation}
It should be noted that $R[\infty]=1$ implies that 
the head of its impulse response is unity. 

We would like to determine the noise shaping filter $R[\mathrm z]$  
that minimizes the MSE under the norm constraint. 
If the amplitude response of the optimal filter can be expressed 
as a rational function, then we can find the noise shaping filter 
that is close to the optimal noise shaping filter. 
Even if this is not the case, we may expect the obtained $R[\mathrm z]$ 
to have a comparable MSE with the optimal noise shaping filter. 

With the state-space expressions of $P[\mathrm z]$ and $R[\mathrm z]$, 
$||P[\mathrm z]R[\mathrm z]||^2$ can be evaluated by 
a bilinear matrix inequality (BMI), whereas
$||R[\mathrm z]||^2$ is evaluated by a 
linear matrix inequality (LMI) \cite{BoydLMIbook}. 
The BMI can be converted into an LMI using a change of variables 
\cite{Masubuchi98, 599969}. 
Thus, as shown in \cite{ohno2016FebSP}, 
the optimization problem is cast into a convex 
optimization problem that can be solved numerically and efficiently 
with a numerical solver such as the CVX\cite{cvx}. 
In this case, the order of $R[\mathrm z]$ should be set to  
be equal to the order of $P[\mathrm z]$ 
because it is the minimum order that can achieve a minimum 
and a higher order for $R[\mathrm z]$ does not reduce the minimum 
\cite{Masubuchi98, 599969}. 

\section{Numerical examples} 
\begin{figure}[t!]
\begin{center}
 \centerline{\includegraphics[width=80mm]{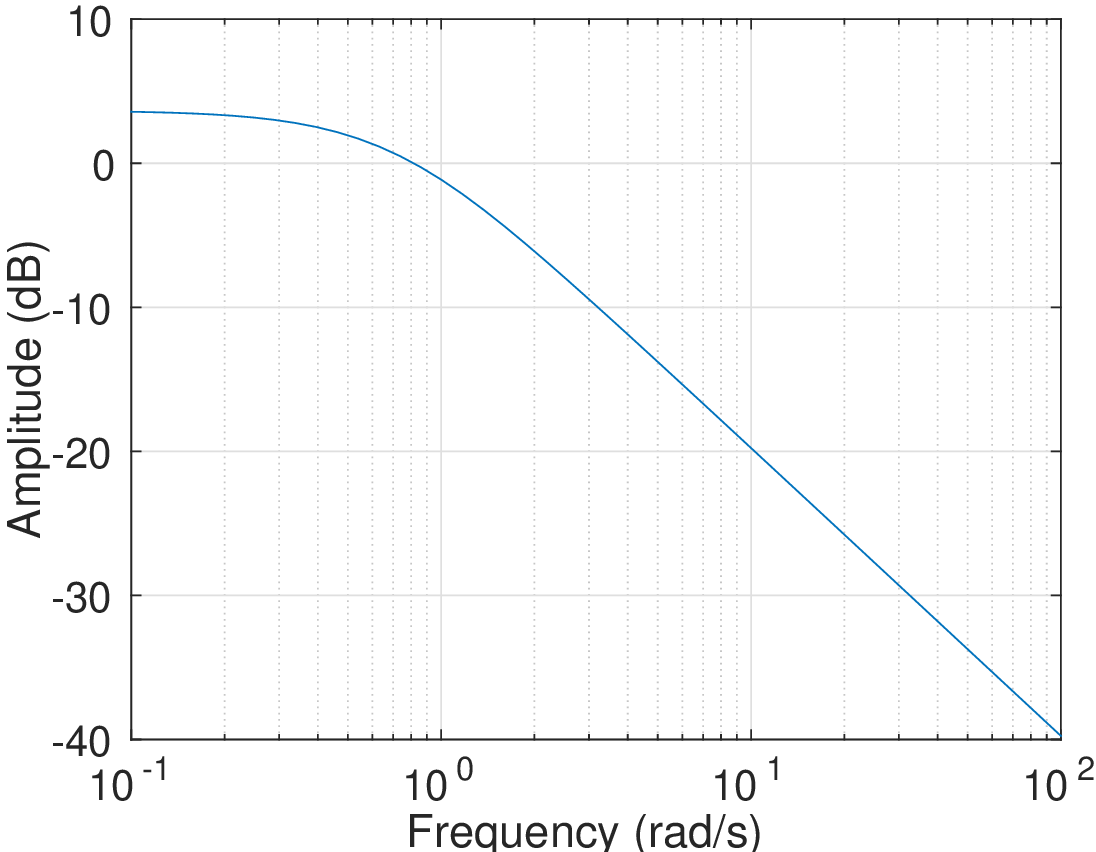}}
 \caption{
Amplitude response of the continuous-time system $P(s)$. 
}
 \label{fig:frq_rsp}
  \end{center}
\end{figure}

To validate our analysis and synthesis, we consider 
a continuous-time system of order four for example  
whose transfer function is  
\begin{equation}
  \label{eq:66}
P(s)=\frac{1.029 s^3 + 4.589 s^2 + 7.146 s + 3.882}
{s^4 + 5.088 s^3 + 9.789 s^2 + 8.296 s + 2.548}
.
\end{equation}
The amplitude response of this system is plotted in Fig. \ref{fig:frq_rsp}. 
We discretize this continuous-time system with 
a sampling period $T_s=0.1$ to obtain the discrete-time 
system $P[\mathrm z]$. 

\begin{figure}[t!]
\begin{center}
 \centerline{\includegraphics[width=80mm]{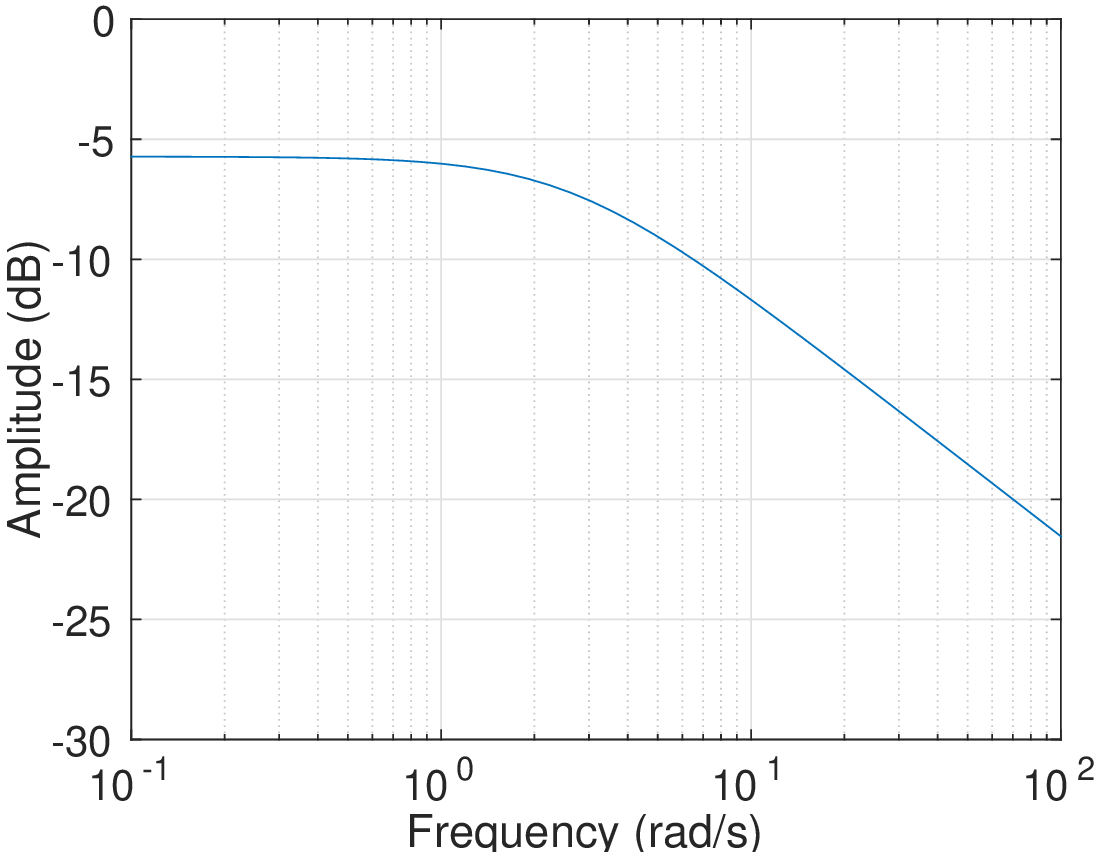}}
 \caption{
Input spectrum. 
}
 \label{fig:frq_omega}
  \end{center}
\end{figure}

We model the continuous-time input signal 
as a stationary process with a zero mean and a spectrum given by  
\begin{equation}
  \label{eq:67}
  S(\omega)=
  c
  \left|
\frac{1}{j\omega+2.62}
  \right|^2
\end{equation}
where $c$ is a constant. 
We set the value of $c$ so that 
the sampled signal should have a unit variance.  
The spectrum is depicted in Fig. \ref{fig:frq_omega}. 

\begin{figure}[t!]
\begin{center}
 \centerline{\includegraphics[width=80mm]{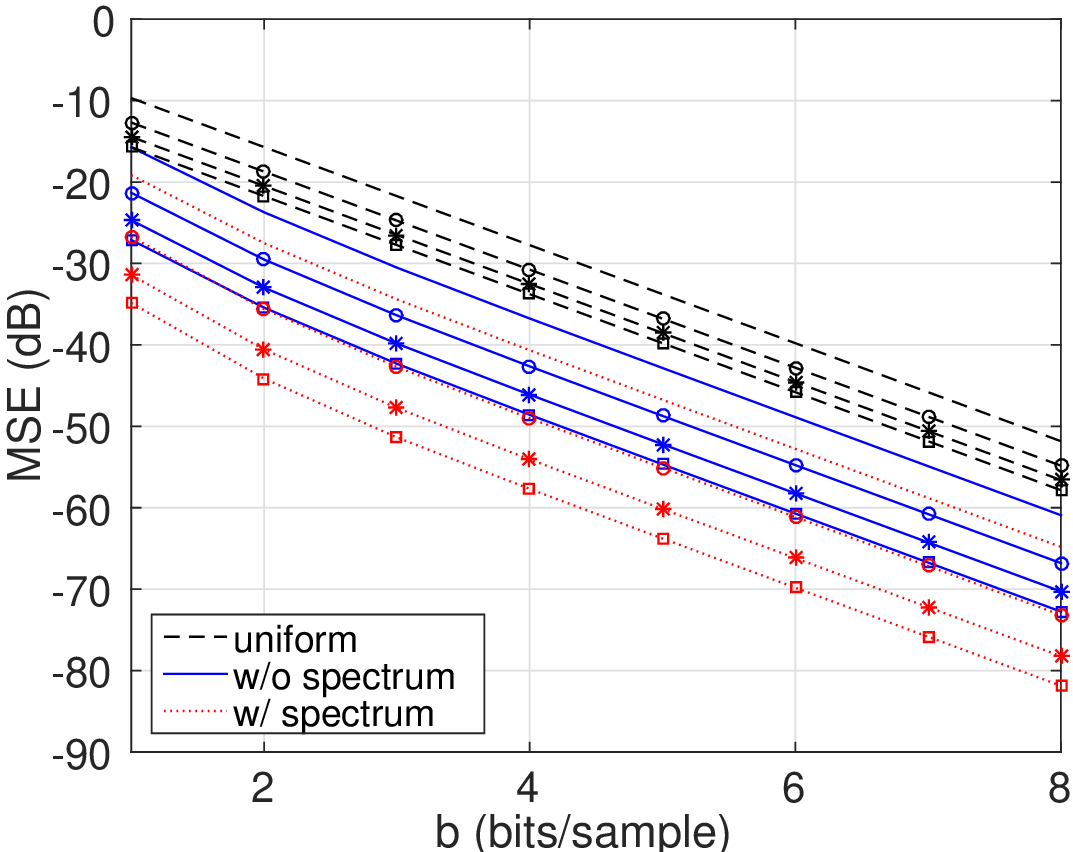}}
 \caption{
MSEs of the optimal feedback quantizer, 
the optimal feedback quantizer \cite{4522532} (dotted curve),  
and the uniform quantizer (dashed curve)  
with different oversampling rates $\lambda$, 
for a colored input, 
where $\circ$, $\ast$, and $\square$ correspond to 
the oversampling ratios $\lambda=2$, $\lambda=3$, 
and $\lambda=4$, respectively. 
}
 \label{fig:bit_color}
  \end{center}
\end{figure}

\begin{figure}[t!]
\begin{center}
 \centerline{\includegraphics[width=80mm]{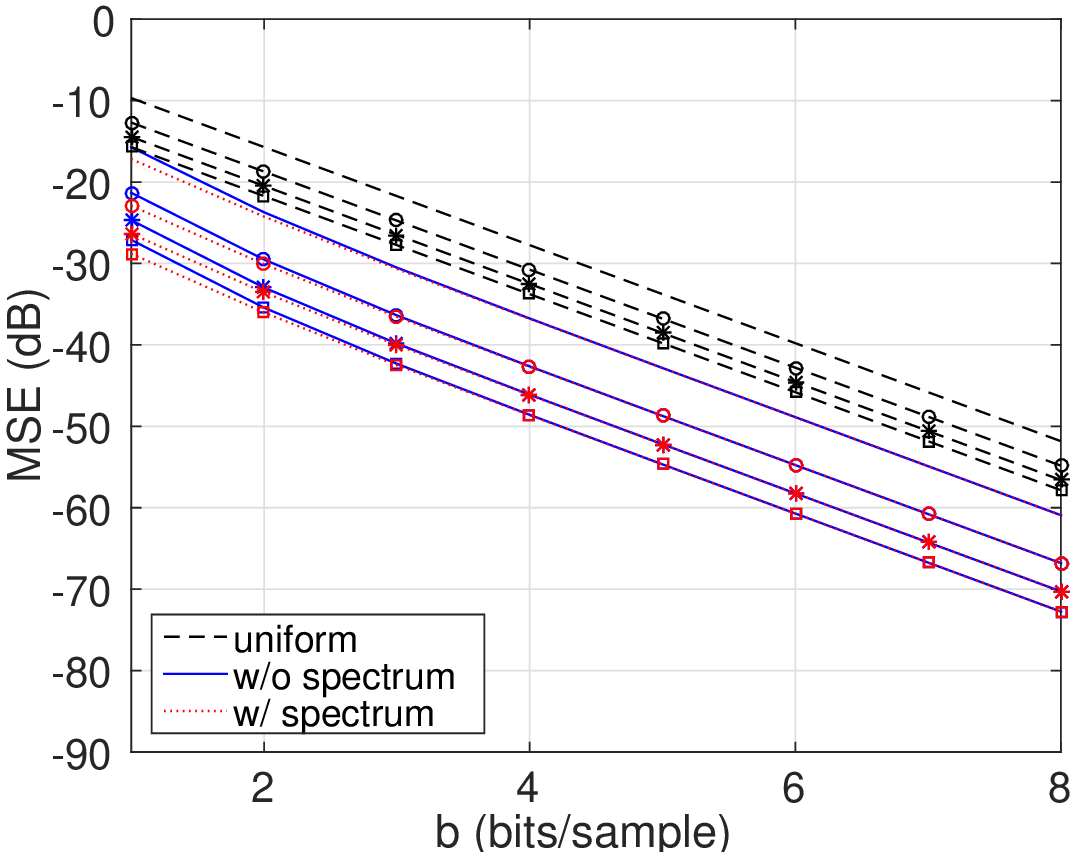}}
 \caption{
MSEs of the optimal feedback quantizer, 
the optimal feedback quantizer \cite{4522532} (dotted curve),  
and the uniform quantizer (dashed curve)  
with different oversampling rates, $\lambda$, 
for a white input, 
where $\circ$, $\ast$, and $\square$ correspond to 
the oversampling ratios $\lambda=2$, $\lambda=3$, 
and $\lambda=4$, respectively. 
}
 \label{fig:bit_white}
  \end{center}
\end{figure}

The loading factor is set to be four. 
For $b=1,2,\ldots,8$, we obtain $\gamma$ from \eqref{eq:63}. 
Then, for a given $\gamma$, we numerically find 
the optimal $\alpha$ from \eqref{eq:23} and \eqref{eq:28}  
that is the minimum MSE (c.f. \eqref{eq:37}),   
replacing $p(\omega)$ by $p_{\lambda}(\omega)$ in \eqref{eq:34}.

For the oversampling ratio $\lambda=1,2,3,4$,  
Fig. \ref{fig:bit_color} compares 
the MSEs of the optimal feedback quantizer,  
the optimal feedback quantizer 
with the pre-/post-filters \cite{4522532} (dotted curve),  
and the uniform quantizer (dashed curve), 
where $\circ$, $\ast$, and $\square$ correspond to 
the oversampling ratios $\lambda=2$, $\lambda=3$, 
and $\lambda=4$, respectively. 

The feedback quantizer has an approximately 10 dB gain against 
the uniform quantizer that is enabled 
by utilizing the feedback filter that is optimized 
based on the system $P[\mathrm z]$. 
A further gain is obtained by 
exploiting the input spectrum 
for the quantizer having 
an optimized feedback filter and pre-/post-filters. 
For all quantizers, as the oversampling ratio increases, 
the MSE decreases and the increment of the MSE gain decreases. 

Fig. \ref{fig:bit_white} shows the MSEs of 
the optimal feedback quantizer, 
the optimal feedback quantizer with the pre-/post-filters  
and the uniform quantizer for a white input signal. 
The optimal feedback quantizer and the optimal feedback quantizer with the pre-/post-filters have a gain of more than 10 dB over the uniform quantizer. 
As the input has a flat spectrum, 
the optimal feedback quantizer has almost the same 
performance as the optimal feedback quantizer with the pre-/post-filters. 
It should be noted that the latter requires additional pre-/post-filters. 
 
In Fig. \ref{fig:bit_color} and Fig. \ref{fig:bit_white}, 
we have utilized ideal feedback filters 
both for the feedback quantizer and the feedback quantizer with the pre-/post-filters, 
which cannot be implemented in practice.   
We approximate the ideal feedback filters for the optimal feedback quantizers 
using IIR filters of order four by the Yule-Walker method 
\cite{4103917} with a normalization 
and by the LMI-based method discussed in Section \ref{sec:design}. 

\begin{figure}[t!]
\begin{center}
 \centerline{\includegraphics[width=80mm]{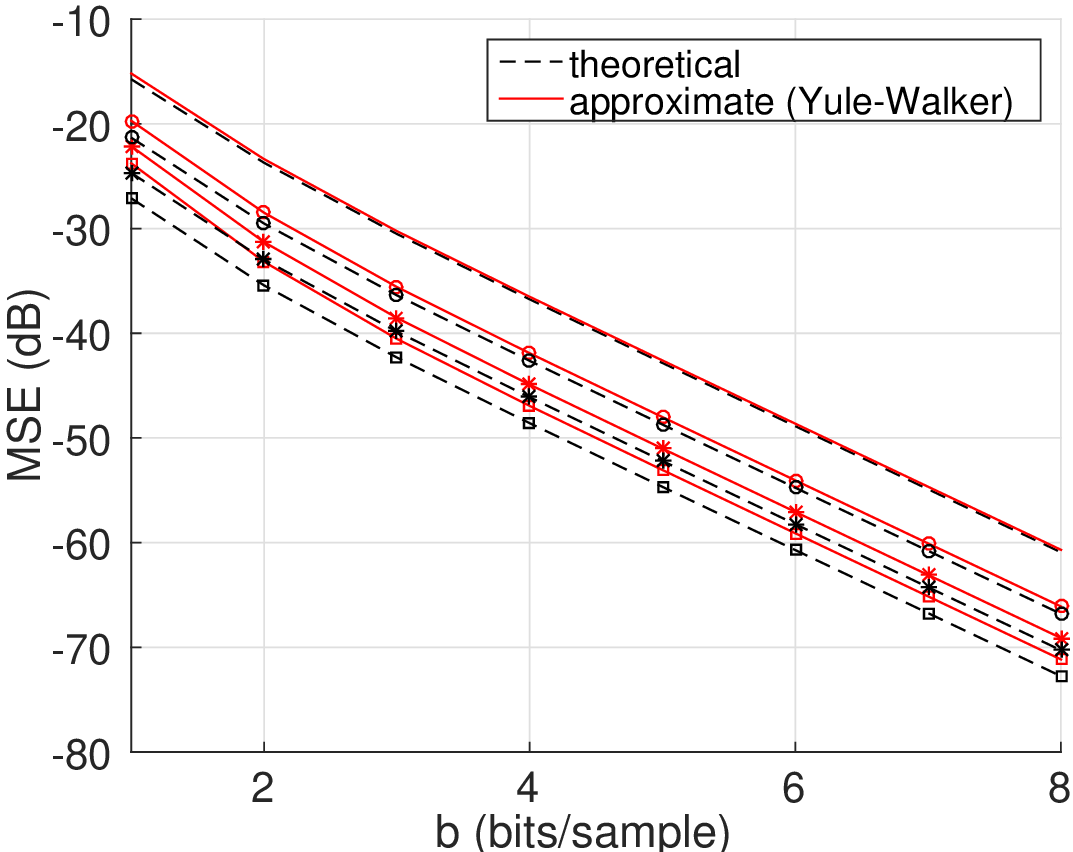}}
 \caption{
MSEs of the feedback quantizers with ideal feedback filters 
and feedback quantizers with 
IIR feedback filters of order four approximated by the Yule-Walter method 
for different oversampling rates $\lambda$, 
where $\circ$, $\ast$, and $\square$ correspond to 
the oversampling ratios $\lambda=2$, $\lambda=3$, 
and $\lambda=4$, respectively. 
}
 \label{fig:YW}
  \end{center}
\begin{center}
 \centerline{\includegraphics[width=80mm]{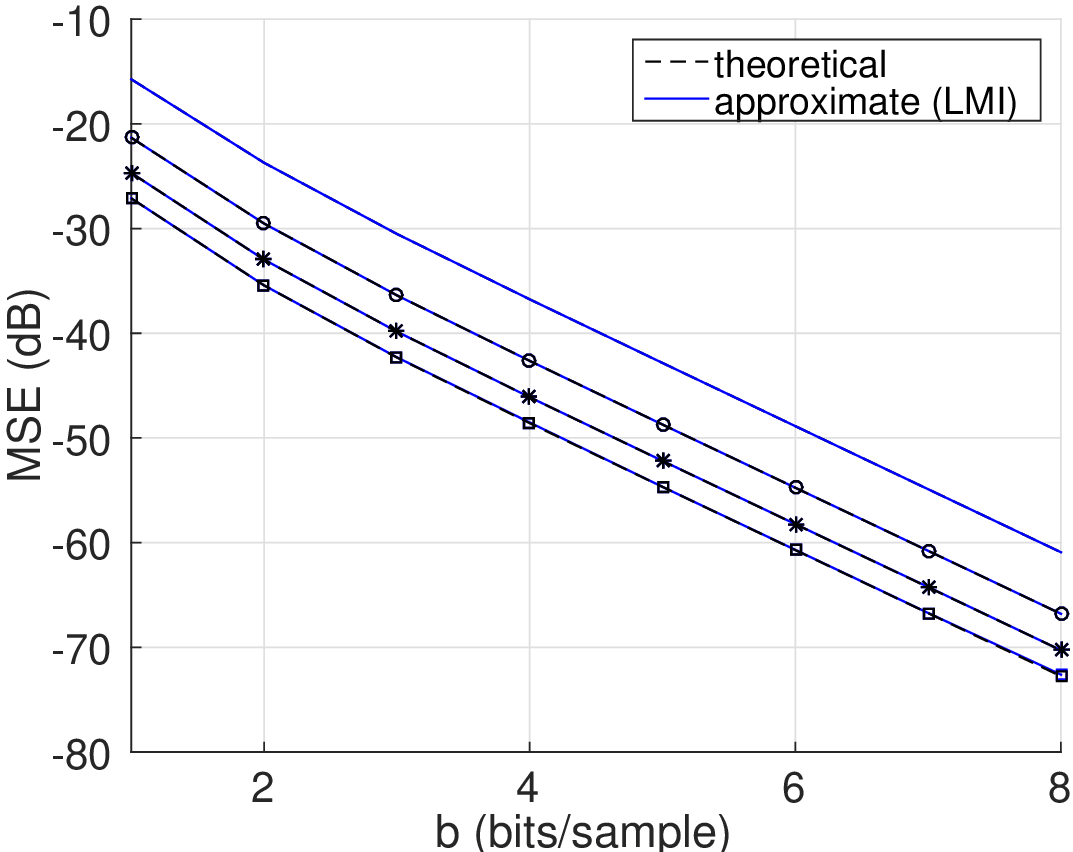}}
 \caption{
MSEs of the feedback quantizers with optimal feedback filters 
and feedback quantizers with 
IIR feedback filters of order four approximated by the LMI-based method 
for different oversampling rates $\lambda$,  
where $\circ$, $\ast$, and $\square$ correspond to 
the oversampling ratios $\lambda=2$, $\lambda=3$, 
and $\lambda=4$, respectively. 
}
 \label{fig:LMI}
  \end{center}
\end{figure}

Fig. \ref{fig:YW} illustrates the MSEs of 
the feedback quantizers with ideal optimal feedback filters  
and 
the feedback quantizers with feedback filters of order four 
approximated by the Yule-Walker method, 
whereas Fig. \ref{fig:LMI} presents the MSEs of  
the feedback quantizers with ideal feedback filters and
the feedback quantizers with feedback filters of order four
 approximated by the LMI-based method.  
The approximation by the Yule-Walker method suffers  
a small loss, while the approximation by the LMI-based method has 
almost the same MSE as the ideal case. 

If the order of the IIR filter is increased, a better performance 
can be expected for the Yule-Walker method. 
On the other hand, it is known that 
the minimum of \eqref{eq:41} is attained 
by $P[\mathrm z]$ having the same order as $R[\mathrm z]$ 
\cite{Masubuchi98, 599969}. 
Therefore, if the order of $P[\mathrm z]$ is increased 
more than the order of $R[\mathrm z]$, the MSE does not improve. 
In this example, 
as the order of $P[\mathrm z]$ is four, an $R[\mathrm z]$ of order four 
is sufficient for the LMI-based method. 
The performance difference 
between the Yule-Walker method and the LMI-based method 
may be decreased by increasing the filter order 
for the Yule-Walker method. 

\section{Conclusions}
We have presented the rate-distortion analysis of 
quantizers with error feedback. 
We have shown that the amplitude response of 
the optimal error feedback filter that minimizes the MSE 
can be parameterized by one parameter and can be found numerically. 
With the optimal error feedback filter, the relationship between 
the number of bits used for the quantization and the achievable MSE 
has been clarified. We have also developed two designs 
for the IIR error feedback filters for approximating the 
ideal optimal error feedback filters.   
Numerical examples have been 
provided to demonstrate our analysis and synthesis. 

\clearpage

\clearpage
\section{Proof of Theorem \ref{theorem:1}}
\label{sec:appendix1}

Suppose that $r(\omega)$ is optimal. 
If $c_0>0$, then $r^{\prime}(\omega)= r(\omega)e^{-c_0}$ gives a smaller 
value for \eqref{eq:18} 
that contradicts the optimally of $r(\omega)$. 
Thus $c_0$ in \eqref{eq:15} has to be zero. 

Let us denote the norm of $r_{opt}(\omega)$ as $c_{opt}$ 
and define the set of $r(\omega) \in {\cal C}_0$ having the same norm as 
$r_{opt}(\omega)$ by ${\cal C}_{opt}$. 
As ${\cal C}_{0} \cap {\cal C}_{opt} \subset {\cal C}_{0} \cap {\cal C}_{1}$, 
the minimization of \eqref{eq:18} subject to ${\cal C}_{0} \cap {\cal C}_{1}$ 
is equivalent to the minimization of $||p(\omega)r(\omega)||^2$ 
subject to 
\begin{align}
  &||r(\omega)||^2=c_{opt}\\
\label{eq:43}
& \frac{1}{2\pi} \int_{-\pi}^{\pi} \ln r(\omega) d \omega =0
\end{align}

The Lagrangian of this problem is given by 
\begin{equation}
  \label{eq:44}
  L(r(\omega)):=p^2(\omega)r^2(\omega)+
\mu_1 r^2(\omega) +\mu_2 \ln r(\omega)
\end{equation}
where $\mu_1$ and $\mu_2$ are the Lagrange multipliers. 
Then, the optimal $r(\omega)$ has to satisfy  
\begin{align}
  \frac{\partial}{\partial r} L(r(\omega))=
2 p^2(\omega)r(\omega)+
2 \mu_1 r(\omega) +\mu_2 \frac{1}{r(\omega)}=0
\quad \mbox{a.e. } \omega \in [-\pi, \pi].   
\end{align}
Thus, for a.e. $\omega \in [-\pi, \pi]$, we need 
\begin{align}
  2(p^2(\omega)+\mu_1)r^2(\omega)=-\mu_2 
.
\end{align}

If $p(\omega)$ is almost constant, then $r(\omega)$ has to be almost constant; 
from \eqref{eq:43} $r(\omega)=1$, implying that 
$R[\mathrm z]=1$. Hence, the error feedback filter 
$R[\mathrm z]-1$ is not required and the uniform quantizer is optimal. 
In the following proof, 
we only consider $p(\omega)$ that is not almost constant.  

As $p(\omega)$ is not almost constant, 
$p^2(\omega)+\mu_1$ cannot be zero 
over any interval $[-\pi, \pi]$, having a nonzero measure.  
As $r(\omega) \ne 0$, $\mu_2$ cannot be zero.    
Therefore, we obtain  
\begin{equation}
  \label{eq:45}
  r(\omega)=\frac{\theta}{\sqrt{p^2(\omega)+\alpha}}
\end{equation}
where $\theta=\sqrt{-\mu_2}$ and $\alpha=\mu_1$. 

Substituting \eqref{eq:45} in \eqref{eq:43} results in  
\begin{align}
  \int_{-\pi}^{\pi} 
  \left(
    \ln \theta -\frac{1}{2}\ln (p^2(\omega)+\alpha) d \omega
  \right)
=0
\end{align}
from which we obtain \eqref{eq:23}. 

\section{Proof of Theorem \ref{theorem:2}}
\label{sec:appendix2}

Differentiating $\Phi(\alpha)$ with respect to $\alpha$, we have  
\begin{equation}
  \label{eq:46}
\dot{\Phi}(\alpha)=\frac{\dot{N}(\alpha)(\gamma+1-C(\alpha))+
N(\alpha)\dot{C}(\alpha)}{[\gamma+1-C(\alpha)]^2}
.
\end{equation}
With \eqref{eq:22}, $N(\alpha)$ can be expressed as  
\begin{equation}
  \label{eq:47}
  N(\alpha)=\frac{\theta^2(\alpha)}{2\pi}
\int_{-\pi}^{\pi} 
\frac{p^2(\omega) }{p^2(\omega)+\alpha} d \omega
.
\end{equation}
From 
\begin{equation}
  \label{eq:48}
  \frac{d}{d \alpha}\theta(\alpha)=
\frac{\theta(\alpha)}{4\pi}
  \int_{-\pi}^{\pi} 
\frac{1}{p^2(\omega)+\alpha} d \omega
\end{equation}
the derivative of $N(\alpha)$ is found to be  
\begin{align}
  \frac{d }{d \alpha}
  N(\alpha)&=
\frac{2\theta(\alpha)}{2\pi}\dot{\theta}(\alpha)
  \int_{-\pi}^{\pi} 
\frac{p^2(\omega)}{p^2(\omega)+\alpha} d \omega
-
\frac{\theta^2(\alpha)}{2\pi}
  \int_{-\pi}^{\pi} 
\frac{p^2(\omega)}{(p^2(\omega)+\alpha)^2} d \omega
\\
\label{eq:49}
&=
\frac{\theta^2(\alpha)}{2\pi}
\left\{
\frac{1}{2\pi}
  \int_{-\pi}^{\pi} 
\frac{1}{p^2(\omega)+\alpha} d \omega
  \int_{-\pi}^{\pi} 
\frac{p^2(\omega)}{p^2(\omega)+\alpha} d \omega
-
  \int_{-\pi}^{\pi} 
\frac{p^2(\omega)}{(p^2(\omega)+\alpha)^2} d \omega
\right\}
\end{align}
It can be seen that $\dot{N}(0)=0$. 
To prove  
\begin{equation}
  \label{eq:50}
  \dot{N}(\alpha) <0 \quad \mbox{for}\quad \alpha <0, \quad 
  \dot{N}(\alpha) >0 \quad \mbox{for}\quad \alpha >0.  
\end{equation}
we introduce the next definition and theorem given in \cite{4522532}. 
\begin{definition}
  We say that two function $\phi$, $\psi$: $[a,b] \rightarrow \mathbb{R}$ 
are similarly functionally related if and only if  
there exists a monotonically increasing function $G(\cdot)$  
such that $\phi=G(\psi)$ for all $x \in [a,b]$.  
Similarly, if there exists a monotonically decreasing 
function such that $\phi=G(\psi)$ for all $x \in [a,b]$, 
we say that $\phi$ and $\psi$ are oppositely functionally related. 
\end{definition}
\begin{theorem}\label{theorem:5}
  If $\phi$, $\psi$: $[a,b] \rightarrow \mathbb{R}$ 
are similarly functionally related, then  
\begin{equation}
  \label{eq:51}
  [b-a]\int_{a}^{b} \phi(x) \psi(x) dx 
\geq 
\int_{a}^{b} \phi(x)dx \int_{a}^{b}\psi(x) dx
. 
\end{equation}
If $\phi$ and $\psi$ are oppositely functionally related, 
then the equality in \eqref{eq:51} is reversed. 
In either case, equality is achieved if and only 
$\psi(x)$ is almost constant. 
\end{theorem}

We set $\psi(\omega)=\frac{1}{p^2(\omega)+\alpha}$ and 
$\phi(\omega)=\frac{p^2(\omega)}{p^2(\omega)+\alpha}$ 
that are related to $\alpha\ne 0$ such that 
\begin{equation}
  \label{eq:52}
\phi(\omega)= \frac{p^2(\omega)}{p^2(\omega)+\alpha}=
1-\frac{\alpha}{p^2(\omega)+\alpha}
=1-\alpha \psi(\omega). 
\end{equation}
Thus, $\phi(\omega)$ and $\psi(\omega)$ are similarly functionally related 
for $\alpha <0$, whereas 
$\phi(\omega)$ and $\psi(\omega)$ are oppositely functionally related 
for $\alpha >0$. 
Then, we can apply theorem \ref{theorem:5} to find that  
\[
 \frac{1}{2\pi} 
  \int_{-\pi}^{\pi} \psi(\omega) d \omega
\int_{-\pi}^{\pi} \phi(\omega) d \omega
-
  \int_{-\pi}^{\pi} \phi(\omega)\psi(\omega) d \omega
\]
is negative for $\alpha <0$, whereas it is positive for $\alpha >0$, 
proving \eqref{eq:50}. 

On the other hand, differentiating $C(\alpha)$ with respect to 
$\alpha$ gives  
\begin{align}
  \frac{d}{d \alpha}C(\alpha)&=
\frac{2\theta(\alpha)}{2\pi}\dot{\theta}(\alpha)
  \int_{-\pi}^{\pi} 
\frac{1}{p^2(\omega)+\alpha} d \omega
-
\frac{\theta^2(\alpha)}{2\pi}
  \int_{-\pi}^{\pi} 
\frac{1}{(p^2(\omega)+\alpha)^2} d \omega
\\ \label{eq:53}
&=
\frac{\theta^2(\alpha)}{2\pi}
\left\{
\frac{1}{2\pi}
\left(
  \int_{-\pi}^{\pi} 
\frac{1}{p^2(\omega)+\alpha} d \omega
\right)^2
-
  \int_{-\pi}^{\pi} 
\frac{1}{(p^2(\omega)+\alpha)^2} d \omega
\right\}
\end{align}
From the Cauchy-Schwarz inequality, 
we find that $\dot{C}(\alpha) < 0$. 

We note that $\gamma+1-C(\alpha)>0$ and $N(\alpha)>0$ in \eqref{eq:46}. 
For $\alpha <0$, from $\dot{N}(\alpha)<0$ and $\dot{C}(\alpha) < 0$ 
in \eqref{eq:46}, $\dot{\Phi}(\alpha)<0$. 
At $\alpha =0$, from $\dot{N}(0)=0$, we have  
\begin{equation}
  \label{eq:54}
  \dot{\Phi}(0)
=\frac{N(0)\dot{C}(0)}{[\gamma+1-C(0)]^2}<0. 
\end{equation}
As $\Phi(\alpha)$ is continuous in $\alpha$, the minimum of 
$\Phi(\alpha)$ is achieved at $\alpha $ greater than zero; 
i.e., we can conclude that $\alpha_{opt} >0$.  

A necessary condition for $\alpha_{opt}$ is $\dot{\Phi}(\alpha_{opt})=0$.  
As $\dot{N}(\alpha) \ne 0$ for $\alpha>0$ and $\alpha_{opt} >0$, 
we find from \eqref{eq:46} that 
the numerator has to be zero, leading to  
\begin{equation}
  \label{eq:55}
  \gamma+1=
\frac{\dot{N}(\alpha_{opt})C(\alpha_{opt})-N(\alpha_{opt})\dot{C}(\alpha_{opt})}{\dot{N}(\alpha_{opt})}
.
\end{equation}

From \eqref{eq:49} and \eqref{eq:53}, we get  
\begin{align}
\nonumber 
& 
\left(
\dot{N}(\alpha_{opt})C(\alpha_{opt})-N(\alpha_{opt})\dot{C}(\alpha_{opt})
\right)/
\left(
\frac{\theta^2(\alpha_{opt})}{2\pi}
\right)^2
\\ \nonumber
=&  
\left\{
\frac{1}{2\pi}
  \int_{-\pi}^{\pi} 
\frac{1}{p^2(\omega)+\alpha_{opt}} d \omega
  \int_{-\pi}^{\pi} 
\frac{p^2(\omega)}{p^2(\omega)+\alpha_{opt}} d \omega
-
  \int_{-\pi}^{\pi} 
\frac{p^2(\omega)}{(p^2(\omega)+\alpha_{opt})^2} d \omega
\right\}
\int_{-\pi}^{\pi} 
\frac{1}{p^2(\omega)+\alpha_{opt}} d \omega
\\ \nonumber
-& 
\int_{-\pi}^{\pi} 
\frac{p^2(\omega) }{p^2(\omega)+\alpha_{opt}} d \omega
\left\{
\frac{1}{2\pi}
\left(
  \int_{-\pi}^{\pi} 
\frac{1}{p^2(\omega)+\alpha_{opt}} d \omega
\right)^2
-
  \int_{-\pi}^{\pi} 
\frac{1}{(p^2(\omega)+\alpha_{opt})^2} d \omega
\right\}
\\ \label{eq:56}
=&
-  \int_{-\pi}^{\pi} 
\frac{p^2(\omega)}{(p^2(\omega)+\alpha_{opt})^2} d \omega
\int_{-\pi}^{\pi} 
\frac{1}{p^2(\omega)+\alpha_{opt}} d \omega
+
\int_{-\pi}^{\pi} 
\frac{p^2(\omega) }{p^2(\omega)+\alpha_{opt}} d \omega
  \int_{-\pi}^{\pi} 
\frac{1}{(p^2(\omega)+\alpha_{opt})^2} d \omega
.
\end{align}
Substituting  
\begin{equation}
  \label{eq:57}
  \frac{1}{p^2(\omega)+\alpha_{opt}}=
\frac{1}{\alpha_{opt}}
  \left(
1-\frac{p^2(\omega)}{p^2(\omega)+\alpha_{opt}}
  \right)
\end{equation}
in \eqref{eq:56} results in  
\begin{align}
\nonumber
&-
  \int_{-\pi}^{\pi} \frac{p^2(\omega)}{(p^2(\omega)+\alpha_{opt})^2} d \omega
\int_{-\pi}^{\pi} 
\frac{1}{\alpha_{opt}}
  \left(
1-\frac{p^2(\omega)}{p^2(\omega)+\alpha_{opt}}
  \right)
d \omega
\\ \nonumber
&+
\int_{-\pi}^{\pi} 
\frac{p^2(\omega) }{p^2(\omega)+\alpha_{opt}} d \omega
  \int_{-\pi}^{\pi} 
\frac{1}{\alpha_{opt}}
  \left(
1-\frac{p^2(\omega)}{p^2(\omega)+\alpha_{opt}}
  \right)
\frac{1}{p^2(\omega)+\alpha_{opt}} d \omega
\\ \nonumber
=&
-\frac{2\pi}{\alpha_{opt}}
  \int_{-\pi}^{\pi} 
\frac{p^2(\omega)}{(p^2(\omega)+\alpha_{opt})^2} d \omega
+\frac{1}{\alpha_{opt}}
\int_{-\pi}^{\pi} 
\frac{p^2(\omega) }{p^2(\omega)+\alpha_{opt}} d \omega
  \int_{-\pi}^{\pi} 
\frac{1}{p^2(\omega)+\alpha_{opt}} d \omega
\\
=&\frac{2\pi}{\alpha_{opt}} \dot{N}(\alpha_{opt})/
\left(
\frac{\theta^2(\alpha_{opt})}{2\pi}
\right)
, 
\end{align}
which shows that  
\begin{equation}
  \label{eq:58}
\dot{N}(\alpha_{opt})C(\alpha_{opt})-N(\alpha_{opt})\dot{C}(\alpha_{opt})=
\frac{\theta^2(\alpha_{opt})}{\alpha_{opt}}\dot{N}(\alpha_{opt}). 
\end{equation}
Substituting this in \eqref{eq:55} gives \eqref{eq:28}. 

\section{Proof of Theorem \ref{theorem:3}}
\label{sec:appendix3}
From \eqref{eq:28}, we obtain  
\begin{align}
  \nu&=\exp
  \left(
    \frac{1}{2\pi} \int_{-\pi}^{\pi} \ln
    \left[
      p_{\lambda}^2(\omega)+
    \alpha_{opt}(\nu,\lambda)
    \right]
    d \omega
  \right)/\alpha_{opt}(\nu,\lambda)
\\
  \label{eq:59}
&=\exp
  \left(
    \frac{1}{2\pi} \int_{-\pi}^{\pi} \ln
    \frac{p_{\lambda}^2(\omega)+  \alpha_{opt}(\nu,\lambda)}{\alpha_{opt}(\nu,\lambda)}
    d \omega
  \right)
.
\end{align}
Substituting \eqref{eq:34} in \eqref{eq:59}, we have 
\begin{align}
  \nu&=\exp
  \left(
    \frac{1}{2\pi} \int_{-\omega_c}^{\omega_c} \ln
    \frac{
      p_{1}^2(\lambda \omega)+  
      \alpha_{opt}(\nu,\lambda)}{\alpha_{opt}(\nu,\lambda)}
    d \omega
  \right)
.
\end{align}
The change of the variable as $\omega^{\prime}=\lambda \omega$ gives  
\begin{equation}
  \label{eq:60}
  \nu=
\exp
  \left(
    \frac{1}{2\pi \lambda} \int_{-\pi}^{\pi} \ln
    \frac{p_{1}^2(\omega^{\prime})+  
      \alpha_{opt}(\nu,\lambda)
    }{\alpha_{opt}(\nu,\lambda)
    }
    d \omega^{\prime}
  \right).  
\end{equation}
Then we have 
\begin{equation}
  \label{eq:61}
  \nu^{\lambda}=
\exp
  \left(
    \frac{1}{2\pi} \int_{-\pi}^{\pi} \ln
    \left[
      p_{1}^2(\omega^{\prime})+  
      \alpha_{opt}(\nu,\lambda)
    \right]
    d \omega^{\prime}
  \right)/\alpha_{opt}(\nu,\lambda)
\end{equation}
that proves 
\begin{equation}
  \label{eq:62}
  \alpha_{opt}(\nu,\lambda)=
\alpha_{opt}(\nu^{\lambda},1)
\end{equation}
hence \eqref{eq:38}. 
\end{document}